# Two-Dimensional Pnictogen Honeycomb Lattice: Structure, On-Site Spin-Orbit Coupling and Spin Polarization

Jason Lee[1], Wen-Chuan Tian[1], Wei-LiangWang[1], Dao-Xin Yao[1*]

Because of its novel physical properties, two-dimensional materials have attracted great attention. From first-principle calculations and vibration frequencies analysis, we predict a new family of two-dimensional materials based on the idea of octet stability: honeycomb lattices of pnictogens (N, P, As, Sb, Bi). The buckled structures of materials come from the sp3 hybridization. These materials have indirect band gap ranging from 0.43 eV to 3.7 eV. From the analysis of projected density of states, we argue that the s and p orbitals together are sufficient to describe the electronic structure under tight-binding model, and the tight-binding parameters are obtained by fitting the band structures to first-principle results. Surprisingly large on-site spin-orbit coupling is found for all the pnictogen lattices except nitrogen. Investigation on the electronic structures of both zigzag and armchair nanoribbons reveals the possible existence of spin-polarized ferromagnetic edge states in some cases, which are rare in one-dimensional systems. These edge states and magnetism may exist under the condition of high vacuum and low temperature. This new family of materials would have promising applications in electronics, optics, sensors, and solar cells.

[1]State Key Laboratory of Optoelectronic Materials and Technologies, School of Physics and Engineering, Sun Yat-Sen University, Guangzhou 510275, China. Correspondence and requests for materials should be addressed to D. X. Y. (email: yaodaox@mail.sysu.edu.cn)

Exploration of unknown phases of materials has been the scientific endeavor for the past decades, and the discovery of new materials can spurt new field of study for both experimentalists and theorists. The low dimensional systems are gaining increasing attention in recent years. Low dimensional systems generally refer to systems in zero-, one- or two-dimension. In particular, various two-dimensional (2D) materials are found in recent years[1,2]. The novel properties of 2D materials make them important in both fundamental research and applications. For example, graphene has remarkable mechanical, thermal, and electronic properties. 2D transition metal dichalcogenides have peculiar optical properties[3,4]. They have promising applications in optoelectronics, spintronics, catalysts, chemical and biological sensors, supercapacitors, and solar cells[5]. In other words, the study of 2D materials is of great importance.

Successful exfoliation of graphene marked the beginning of the study of 2D materials[6,7,8]. However, the application of graphene is limited by the lack of intrinsic gap. The planer structure of graphene also renders it impossible to control the electronic structure by perpendicular electric field. Recently, black phosphorus was synthesized and studied[9,10,11], which has a direct band gap and high carrier mobility[8]. Its puckered structure also enables the modification of band structure by perpendicular field. Recently, another allotrope of phosphorus called blue phosphorus was predicted. It has buckled structure similar as silicene and germanene, and its band gap is close to black phosphorus[12]. Afterwards, two additional phosphorus allotropes were predicted by first-principle study[13]. Accordingly, the phosphorus allotropes show great importance in both theory and experiment.

Phosphorus is pnictogen, and belongs to the nitrogen group (VA). Besides the study of phosphorus, 2D systems involving other VA elements are also carried out. Quantum spin hall effect is predicted to exist in honeycomb X-hydride/halide (X=N-Bi) monolayers[14] and BiX/SbX (X=H, F, Cl and Br) monolayers[15]. Arsenic, antimony and bismuth honeycomb monolayers are also studied in topological aspects, and they are found to be topological insulators[16,17,18,19].

Despite all these studies, it is still unknown whether all pnictogens can have honeycomb lattice structure, especially nitrogen. Secondly, the orbitals involved in the electronic structures near Fermi level are unclear, thus no tight-binding models applicable to these systems have been proposed. The spin-orbit couplings (SOC) are believed to play important role in the electronic structure, which have not been studied seriously. Furthermore, the properties of nanoribbons based on these materials are not known well, such as their magnetism. These questions are highly nontrivial and worth systematic study.

In this paper, we use first-principle calculation to systematically explore the unknown phase of VA elements in the periodic table. Both the structures of bulk and 2D monolayers are optimized and found to be stable. The electronic structures of monolayers are intensively studied, and the evolution of band structures is clearly shown. A tight-binding Hamiltonian is constructed, and the hopping parameters are obtained by fitting to band structures from first-principle calculations. We find large SOC for all elements except nitrogen, in which the on-site contribution is the leading term. Unlike other types of SOC, on-site SOC is solely determined by atomic orbitals of lattice sites; and on-site SOC does not break time-reversal symmetry, therefore these materials can be good candidate for topological superconductors. Electronic structures of both armchair and zigzag nanoribbons of these materials are also studied. The conducting edge states appear in nitrogen and bismuth zigzag nanoribbon. More interestingly, ferromagnetism shows up in the zigzag nanoribbons of P, As and Sb, and in the armchair nanoribbons of As and Sb. The ferromagnetic edge states may contribute to useful applications, such as spin-polarized angular-resolved photoemission spectroscopy (ARPES), spin-polarized scanning tunneling microscopy (SP-STM), and spin-polarized field emission. On the other hand, the appearance of conducting edge states on nitrogen zigzag nanoribbon is a novel and counter-intuitive phenomenon that worth deep investigation.

## Results

**Layered Structure and Monolayers**
The three-dimensional (3D) AB stacking structures and corresponding 2D structures of VA elements are optimized by first-principle calculations. We calculated vibration frequencies for monolayers of all VA elements, and no imaginary frequency was found, suggesting the stability of all these structures. Pnictogens have five outer electrons, and they need three additional electrons to achieve octet stability. Their hybridized $sp^3$ orbitals account for their similar buckled structures, as shown in Figure 1, and the optimized structure parameters of nitrogen (N), phosphorus (P), arsenic (As), antimony (Sb) and bismuth (Bi) with and without van de Waals (vdW) correction, are listed in Table 1. We use $l_{bond}$ to denote bond length, and $l'$ to denote nearest distance between atoms in adjacent layers. $d$ and $\Delta z$ stand for inter-layer distance and buckling. $a$ is the lattice constant. As shown in Figure 1, each layer in these 3D layered structures, has silicene-like structure. The calculated $a$ and $l_{bond}$ for P are 3.32Å/3.28 Å (with/without vdW) and 2.27 Å for bulk, respectively, and 3.28 Å and 2.27 Å for monolayer, respectively, in agreement with previous study by Z. Zhu *et.al.*[7], although their inter-layer distance is larger than our value.

As shown in Table 1, the $l_{bond}$'s are much smaller than the $l'$'s. The structure parameters of monolayer differ only slightly from that of the bulk. It is clear that the structure parameters for nitrogen are much smaller than those of other VA elements. The inclusion of the vdW correction mainly affects $d$ while having only minor effect on the $a$'s, $\Delta z$'s and $l_{bond}$'s. When vdW is included, all the $l'$'s are about twice the vdW radii. In other words, the inclusion of vdW interaction does not dramatically change the structure parameters within a layer, even though there are noticeable changes in the $d$'s, such as for P and As. All these phenomena suggest that layers couple with each other by weak vdW interaction and thus can easily exfoliated to form monolayer [20]. This theoretical prediction may provide guidance to experimental synthesis and applications.

Even though there is no monotonous increase of $d$'s, the monotonous increases of $a$, $\Delta z$'s, $l_{bond}$'s and $l'$'s with respect to increasing atomic number from N to Bi are clear. This is in agreement with the periodic law of elements, in that the atoms become larger with increasing atomic number. Interestingly, if we compare the bond length with twice the covalent radius, the bond length is a little bit longer.

It is necessary to point out that, during the optimization process of three-dimensional (3D) structure of N, we found the tiny jittering of total energy as the inter-layer distance changes. This may suggest that the interaction between N layers are extraordinarily weak (if they have any), and the calculations find it difficult to optimize the structure to minimize the total energy. This means that these layers can detach from each other very easily, therefore the 3D structure of N may not exists, while the 2D structure of N can exist. Moreover, due to the interaction between the sheet and the substrate, the presence of substrate can enhance the stability of the monolayer. Experimentally, the 2D nitrogen atomic sheet has been grown epitaxially on GaAs(001) surface[21].

**Electronic Structure of Monolayer**
The band structures obtained from our first-principle calculations shown in Figure 2 (without SOC) and Figure 3 (with SOC), and the band gaps for all the monolayers are listed in Table 3, with and without SOC. The projected density of states (PDOS) are presented in Figure 4.

From the analysis of PDOS, it is clear that the electronic structures of all the monolayers are mainly made up of s and p orbitals. N has no 2d orbital; P has 3d orbitals, but they are vacant. Therefore, in the case of N and P monolayer, the contribution of the d orbitals to the total density of states (DOS) is exactly zero. For As, Sb and Bi, SOC in d orbitals has non-zero but only minor contribution to the total DOS. Based on this analysis, we argue that the band structures of all these monolayers can be depicted by s and p orbitals, and d orbitals are not necessary. Meanwhile, the gap between the lower branch and the middle branch enlarges as atomic number increases due to the relativistic effect.

From the band structures shown in Figure 2 and Figure 3, three branches of bands are evident, and they are represented by three different colors in the figures. For all the cases, and the lower branch mainly consists of s orbitals while the other two branches are mainly from p orbitals. For N monolayers, the band structure resembles that of silicene[22]; PDOS shows that the lower branch and the middle branch are entangled. As atomic number increases, the lower branch gradually disentangles from the middle one, becoming lower in energy, and formed a graphene-like band structure. From analysis of PDOS presented in Figure 4, the s component of the lower branch increases with decreasing p character, but the middle and upper branches show an opposite trend: the p character increases as s character decreases. It can be seen from both the band structures and PDOS's, the ranges of all these three bands shrink, indicating the weakening of interactions between orbitals as the covalent bonds become longer.

As shown in Table 3, the band gap decreases from 3.70eV of nitrogen to just 0.55eV/0.42eV(without/with SOC) of Bi as atomic number increases, representing a transition from insulator to semiconductor. This is consistent with the periodic law of element: the metal characteristic increases as atomic number increases in VA elements. We obtained a slightly smaller band gap (1.97eV) than previous study of blue phosphorus monolayer (~2eV)[9]. In the absence of SOC, the Bi monolayer has direct band gap while all others have indirect band gaps. The valence band maximum (VBM) of As, Sb and Bi, and the conducting band minimum (CBM) of Bi are located in $\Gamma$ point while others are not. When SOC is included, all band structures have indirect band gaps, and the VBM of Bi shifts away from $\Gamma$ point. We also apply strain to the bismuth monolayer, and find that the band gap decreases with increasing the tensile strain, then it closes at approximately 5% strain, and reopens if strain keeps increasing. This is consistent with Chuang's studies that focus on the topological aspect of bismuth monolayer[19]; they claimed that bismuth monolayer is non-trivial 2D topological insulator.

By comparing the band structure with and without SOC, as shown in Figure 2 and Figure 3, respectively, SOC breaks energy degeneracy of certain points in the band structures. As mentioned before, the lower branch is mainly consisted of s orbitals, while the other two branches are mainly from p orbitals, thus SOC has nonzero contribution to p orbitals, but vanishes for s orbitals because the angular moment of s orbital is zero. Therefore, SOC influences the middle and upper branches of bands while having no effect on the lower branch.

**Tight-Binding Approach to Monolayer**
Based on the analysis of PDOS, it is clear that a tight-binding (TB) model with just s and p orbitals is sufficient to describe the system. Recently Zolymoi *et. al.* have constructed a TB model for silicane and germanane with only s and p orbitals[23]. Similarly, we construct a TB model with SOC.

$$\mathcal{H} = \mathcal{H}_0 + \mathcal{H}_1 + \mathcal{H}_2 + \mathcal{H}_{SO} \tag{1}$$

$$\mathcal{H}_0 = \sum_i \left[ \varepsilon_s a^\dagger(R_i) a(R_i) + \varepsilon_p \boldsymbol{b}^\dagger(R_i) \cdot \boldsymbol{b}(R_i) \right] \tag{2}$$

$$\mathcal{H}_1 = \sum_{\langle i,j \rangle} \left[ h_s(i,j) + h_p(i,j) + h_{sp}(i,j) + h.c. \right] \tag{3}$$

$$\mathcal{H}_2 = \sum_{\langle\langle i,j \rangle\rangle} \left[ h'_s(i,j) + h'_p(i,j) + h'_{sp}(i,j) + h.c. \right] \tag{4}$$

$$\mathcal{H}_{SO} = i\lambda_I \sum_{\langle\langle i,j\rangle\rangle \alpha\beta} v_{ij} \sigma^z_{\alpha\beta} b^\dagger_\alpha(\mathbf{R}_i) \cdot b_\beta(\mathbf{R}_j)$$

$$+ i\lambda_R \sum_{\langle\langle i,j\rangle\rangle \alpha\beta} \mu_{ij} (e_{ij} \times \sigma)^z b^\dagger_\alpha(\mathbf{R}_i) \cdot b_\beta(\mathbf{R}_j) + \mathcal{H}_{On-Site} \qquad (5)$$

$$h_s(i,j) = \gamma_{ss} a^\dagger(\mathbf{R}_i) a(\mathbf{R}_j) \qquad (6)$$

$$h_p(i,j) = \gamma_{pp\sigma} [e_{ij} \cdot b^\dagger(\mathbf{R}_i)][e_{ij} \cdot b(\mathbf{R}_j)]$$

$$+ \gamma_{pp\pi} \{b^\dagger(\mathbf{R}_i) \cdot b(\mathbf{R}_j) - [e_{ij} \cdot b^\dagger(\mathbf{R}_i)][e_{ij} \cdot b(\mathbf{R}_j)]\} \qquad (7)$$

$$h_{sp}(i,j) = \gamma_{sp} a^\dagger(\mathbf{R}_i) (e_{ij} \cdot b(\mathbf{R}_j)) \qquad (8)$$

$$h'_s(i,j) = \gamma'_{ss} a^\dagger(\mathbf{R}_i) a(\mathbf{R}_j) \qquad (9)$$

$$h'_p(i,j) = \gamma'_{pp\sigma} [e_{ij} \cdot b^\dagger(\mathbf{R}_i)][e_{ij} \cdot b(\mathbf{R}_j)]$$

$$+ \gamma'_{pp\pi} \{b^\dagger(\mathbf{R}_i) \cdot b(\mathbf{R}_j) - [e_{ij} \cdot b^\dagger(\mathbf{R}_i)][e_{ij} \cdot b(\mathbf{R}_j)]\} \qquad (10)$$

$$h'_{sp}(i,j) = \gamma'_{sp} a^\dagger(\mathbf{R}_i) (e_{ij} \cdot b(\mathbf{R}_j)) \qquad (11)$$

$$e_{ij} = \frac{\mathbf{R}_i - \mathbf{R}_j}{|\mathbf{R}_i - \mathbf{R}_j|}, \quad v_{ij} = \frac{(d_i \times d_j)^z}{|d_i \times d_j|} \qquad (12)$$

Here, $a^\dagger(\mathbf{R}_i)$ and $a(\mathbf{R}_i)$ are the creation and annihilation operators of s electron in lattice site $\mathbf{R}_i$, and $b^\dagger(\mathbf{R}_i) = (b^\dagger_x(\mathbf{R}_i), b^\dagger_y(\mathbf{R}_i), b^\dagger_z(\mathbf{R}_i))$ and $b(\mathbf{R}_i) = (b_x(\mathbf{R}_i), b_y(\mathbf{R}_i), b_z(\mathbf{R}_i))$ are the creation and annihilation operators of p electron in lattice site $\mathbf{R}_i$. $\varepsilon_s$ and $\varepsilon_p$ in $\mathcal{H}_0$ are the on-site energies of the s and p orbitals. $\mathcal{H}_1$ contains all the nearest neighbor hoppings and $\mathcal{H}_2$ contains all the next nearest neighbor hoppings. Both $\mathcal{H}_1$ and $\mathcal{H}_2$ can be divided into three parts: hoppings between s orbitals ($h_s$ and $h'_s$), hoppings between p orbitals ($h_p$ and $h'_p$), and hoppings between s orbital and p orbital ($h_{sp}$ and $h'_{sp}$). $\gamma_{ss}$, $\gamma'_{ss}$, $\gamma_{pp\sigma}$ and others are the corresponding hopping integrals. The first term in $\mathcal{H}_{SO}$ represents intrinsic SOC and the second term is Rashba SOC [24]. $\alpha$ and $\beta$ in $\mathcal{H}_{SO}$ are spin indices. $(d_i \times d_j)^z$ corresponds to the z component of $d_i \times d_j$. $\mu_{ij} = \pm 1$ for A(B) sites. $d_i$ and $d_j$ are the two nearest bonds connecting the next nearest neighbors. $\mathcal{H}_{On-Site}$ represents the on-site SOC that comes from relativistic effect, and it serves as a correction term for the non-relativistic approximation of Dirac equation in central field. It can be proven that $\mathbf{S} \cdot \mathbf{L}$ does not commute with $\mathcal{H}$, and the non-zero matrix elements of on-site SOC between orbitals and spin components are listed in Table 2.

All together there are 13 parameters in this TB Hamiltonian, including three parameters of SOC. We have obtained these parameters by fitting TB band structure to the result from first-principle calculations, as shown in Table 4. The comparison between first-principle band structures and TB band structures are presented in Figure 5 (without SOC) and Figure 6 (with SOC). It can be seen from Table 4 that these parameters obey the periodic law of element, with some abnormality. $\gamma_{sp}$ and $\gamma_{pp\sigma}$ are dominant over all other parameters because of their better overlap between orbitals. The hoppings between nearest neighbors are stronger than those between next nearest neighbors. Since the SOC in N monolayer is negligibly small, the inclusion of SOC has no detectable effect on the band structure, as can be seen from first-principle results. Therefore SOC is not included in the TB model of N monolayer. The contributions of three types of SOC are very different, and on-site SOC plays a more important role than the other two. The on-site SOCs are surprisingly large for P, As, Sb and Bi monolayers. It is clear that the on-site SOC parameter increases with the atomic number, which is consistent with our first-principle calculations. On-site SOC preserves time-reversal invariant, and may play an important role in time-reversal-invariant topological superconductors[25].

**Zigzag and Armchair Nanoribbon**

To be more comprehensive, we investigate nano-ribbons and their properties. Nano-ribbon demonstrate novel physical phenomenon vastly different from the 2D counterpart. We have calculated zigzag nanoribbons (ZNR) and armchair nanoribbons (ANR) with various widths. Figure 7 (a) shows a particular case of ZNR and ANR of 20 atoms wide. Before the calculation of electronic structures, we relax the outermost three atoms on each side. As shown in Figure 7 (b) to (f), the edges of nitrogen and bismuth ZNR, and the edges of nitrogen, phosphorus and bismuth ANR are found to be rearranged, while for other cases only minute atom displacements are observed. The nitrogen atoms on the edge of nitrogen ZNR and ANR moves towards the central plane (Figure 7 (b) and (d)). The edges of bismuth ZNR are obviously bent(Figure 7 (c)).The outermost bond of phosphorus and bismuth ANR are rotated, making the buckling on the edges larger than that in the center of ANR(Figure 7 (e) and (f)). There is even a possible reconstruction of bismuth ANR on the edges in such a way that each atom is bonded to three neighboring ones. The dangling bonds exists on the outermost atoms on the edges, and they are chemically active presence of air. But they can be stable under ideal conditions, such as ultra-vacuum and low temperature. Recently, single dangling bond on Si surface has been observed[26].

Compared with the band structure of monolayer, the correspondence of the branches of bands are obvious. A particular example of phosphorus is shown in Figure 8.The lower branch of ZNR and ANR mimics the counterpart of graphene. Any bands of ZNR and ANR with this correspondence can be obtained by this projection. And the 2D band gap is 2.05eV for phosphorus ZNR, slightly larger than that of the monolayer.

It is interesting to notice that there are two bands of phosphorus monolayer that do not correspond to any of the bands in the monolayer. The two bands in ZNR are doubly degenerate in energy and spin, but their spins are opposite to each other. There is a gap between these two bands, which is about 0.28eV, and the Fermi level lies inside the gap. In other words, this system is ferromagnetic, and demonstrates a magnetic moment about $2.02\mu_B$ per supercell. It is reasonable since the band below Fermi level is doubly degenerate, and is filled with two electrons from each unit cell. This is different from the antiferromagnetic edge state of graphene ZNR[27,28]. On the other hand, the two bands in ANR are four-fold degenerate, but are not spin-polarized. In other words, the edges of ANR of phosphorus are anti-ferromagnetic, possessing a zero magnetic moment. We also calculate the band structures of phosphorus ZNR and ANR that are passivated by hydrogen, and the edge states disappear. Therefore it is possible that dangling bonds are responsible for edge states. One possible explanation for the magnetism is that, the electrons in the dangling bonds are slightly mixed with other states inside the ribbon, and there are slightly delocalized and thus can overlap with the wavefunctions of the dangling bonds on the other edge.

Similar band structures can be obtained for other cases, as shown in Figure 9. Surprisingly, the edge states of N and Bi ZNR, and the edge states of N and P ANR are not spin-polarized. And there seems to be no edge states for Bi ANR. Interestingly, the edge states of N ZNR are conducting, which is in contrast with our common belief that most materials with nitrogen are unlikely to be conductors. The total magnetic moments per unit cell are $2.02\mu_B$, $2.00\mu_B$ and $2.04\mu_B$ for phosphorus, arsenic, and antimony ZNR with 20 atoms wide, respectively, and $4.05\mu_B$ and $3.95\mu_B$ for arsenic and antimony ANR with 20 atoms wide, respectively. And there is no magnetic moment for nitrogen and bismuth ZNR, or nitrogen, phosphorus and bismuth ANR, because their edge states are not spin-polarized. Furthermore, we tried various widths, such as 22, 24, 26, 28 and 30 atoms wide. Their band structures are very similar, showing that the edge states and magnetic moments are not sensitive to the width of ZNR. This may be another indication that the edge states come from dangling bonds.

Analysis of charge density reveals the existence of edge states. We found that the charges of this band are mainly distributed on both edges of ZNR. Figure 10 shows the charge distribution of edge states of phosphorus ZNR with 20 atoms wide. The charge distributions for other elements with various widths are very similar, and are not presented here. Interestingly, for bismuth ANR, it shows no edge states because of possible reconstruction in which no dangling bonds exists that contribute to edge states.

The spin-polarized edge states of ZNR are very useful in applications. It can be used in spin-polarized angular-resolved photoemission spectroscopy (ARPES), a necessary technique in the study of topological insulators. It can also be used in spin-polarized scanning tunneling microscopy (SP-STM) as a tool to detect the spin texture of material surfaces. Likewise, spin-polarized field emission can take advantage of this polarized spin as well. The graphene ZNR may not be useful in all these cases, since the edge state of graphene ZNR is not spin-polarized, as mentioned above.

**Discussion**

Using first-principle calculations, we demonstrate the existence of two-dimensional honeycomb monolayer of VA elements, and determine the corresponding structure parameter. From the three-dimensional structure, it is clear that the systems have AB stacking layered structure (with some uncertainty about N), and each layer is similar to silicene. The geometry of monolayer is very similar to that of the bulk, indicating that layers are bound together by weak van de Waals interaction. Bond length, lattice constant, and buckling increase monotonously with atomic number from nitrogen to bismuth, obeying the periodic law of elements.

We find that the electronic structure obeys the periodic law of elements, as atomic number goes from nitrogen to bismuth. Three branches of bands are obvious in the band structure, and they are similar for all VA elements with minor variations. Spin-orbit coupling (SOC) plays important role in the band structures, and its effect becomes stronger as atomic number increases, however its influence is still limited. It is shown that SOC influences p orbitals, which can be seen from the changes of bands with major characters of p orbitals. There is no noticeable effect of SOC on s orbitals. The band gap decreases monotonously, from 3.7eV for nitrogen to only 0.55eV (without SOC)/0.43eV(with SOC).

From projected density of states, it is clear that the s and p characters of bands changes monotonously. When the atomic number is small, s and p orbitals are mixed, such as for cases of nitrogen and phosphorus. As the atomic number increases, the s character shifts downwards in energy, and becomes the dominant component of lower branch of bands; on the other hand, the p character shifts upwards in energy, becoming the dominant component of the middle branch and upper branches of bands. In all these cases, the effect of d orbital to band structure is negligible. Based on this analysis, we have constructed a tight-binding (TB) Hamiltonian using only one s orbital and three p orbitals from each atom in the unit cell, and the hopping parameters are obtained by fitting the band structures. Large on-site SOCs are found for the monolayers of all these elements except nitrogen

Lastly, we study the atomic and electronic structure of zigzag nanoribbons (ZNR) and armchair nanoribbons (ANR). Rearrangements in nitrogen and bismuth ZNR and nitrogen, phosphorus and bismuth ANR are found, and there is possible reconstruction in bismuth ANR. In the study of electronic structure, we found an obvious correspondence of band structure to that of monolayer. Band structures of ZNR and ANR show spin-polarized ferromagnetic edge states for phosphorus, arsenic and antimony ZNR and arsenic and antimony ANR. The conducting edge states of nitrogen ZNR is also unusual since it contradicts our convention that materials made from nitrogen are usually insulators.

All these findings contribute to the study of future use of these materials. Firstly, these materials can be used in electronic devices. Secondly, because of their large on-site SOC, they are candidates for topological superconductors. Thirdly, the spin-polarized edge states may be useful in spin-polarized angular-resolved photoemission spectroscopy (ARPES), Spin-polarized scanning tunneling microscopy (SP-STM) and spin-polarized field emission.

## Methods

In this paper, we use first-principle calculations to explore the unknown phase of VA elements in the periodic table. Our calculations are based on Plane Augmented Wave (PAW) with Perdew-Burke-Ernzerh (PBE) of exchange-correlation[29] as implemented in the Vienna *Ab initio* Simulation Package (VASP) code[30]. To ensure reliable result, we use large cut-off energy （400eV） for bulk and monolayers, however we use default cut-off energy for nanoribbons. The systems are restricted to periodic boundary conditions. A vacuum at least 15 Å thick is inserted to eliminate the interaction between different units in the study of system with low dimensions (monolayer and nanoribbon). For optimization, ions are relaxed using conjugate-gradient algorithm until the total force is less than 0.01eV/ Å. Grimme's scheme for van de Waals (vdW) correction[31] is used to optimize bulk structure, and results are compared with those obtained without vdW correction. Unfortunately, vdW correction is not included in the case of bismuth because of lack of parameters in Grimme's scheme. We believe that vdW has no substantial effect on the layer structure of bismuth monolayer. Some vdW scheme with higher precision[32] could have been used, but getting accurate vdW energy is not the main interest of this paper, and a simple time-saving scheme of Grimme is sufficient to prove our statements. Spin-orbit coupling (SOC) is included only in the calculation of band structure of monolayers.

## Acknowledgments

The authors thank F. C. Chuang for helpful discussions. This project is supported by National Basic Research Program of China (2012CB821400), NSFC-11074310, NSFC-11275279, Fundamental Research Funds for the Central Universities of China, Specialized Research Fund for the Doctoral Program of Higher Education (20110171110026), and NCET-11-0547.

## Author contributions

J. L. took charge of all the first-principle calculations and analysis of these calculations. D. X. Y., J. L. and W. C. T. proposed the tight-binding Hamiltonian, and all the parameters in this Hamiltonian are fitted to first-principle results by W. C. T.. W. L. W. was in charge of commercial software usage. J. L. and D. X. Y. wrote the paper and all authors commented on it. D. X. Y. was the leader of this study.

# Additional information

Competing financial interests: The authors declare no competing financial interests.

| Table 1 Structure Parameter for Monolayer of VA Elements | | | | | | |
|---|---|---|---|---|---|---|
| Element | | N | P | As | Sb | Bi |
| Bulk (without vdW) | $a$ | 2.29 | 3.28 | 3.62 | 4.21 | 4.46 |
| | $\Delta z$ | 0.70 | 1.24 | 1.40 | 1.62 | 1.71 |
| | $d$ | 3.11 | 4.10 | 3.96 | 2.85 | 2.71 |
| | $l'$ | 3.38 | 4.52 | 4.47 | 3.75 | 3.74 |
| | $l_{bond}$ | 1.50 | 2.27 | 2.51 | 2.92 | 3.09 |
| Bulk (with vdW) | $a$ | 2.29 | 3.32 | 3.65 | 4.11 | - |
| | $\Delta z$ | 0.70 | 1.21 | 1.37 | 1.64 | - |
| | $d$ | 2.76 | 2.88 | 2.82 | 2.84 | - |
| | $l'$ | 3.06 | 3.46 | 3.52 | 3.70 | - |
| | $l_{bond}$ | 1.50 | 2.27 | 2.51 | 2.89 | - |
| Monolayer | $a$ | 2.30 | 3.28 | 3.61 | 4.12 | 4.33 |
| | $\Delta z$ | 0.70 | 1.24 | 1.40 | 1.65 | 1.71 |
| | $l_{bond}$ | 1.50 | 2.26 | 2.51 | 2.89 | 3.09 |
| Covalent radius[32] | | 0.71 | 1.07 | 1.19 | 1.39 | 1.48 |
| vdW radius[32] | | 1.55 | 1.80 | 1.85 | 2.06 | 2.07 |
| Unit in Å | | | | | | |

| Table 2 Non-Zero Matrix Elements of On-Site SOC | | | | | | | |
|---|---|---|---|---|---|---|---|
| | | spin-up | | | spin-down | | |
| | | $p_x$ | $p_y$ | $p_z$ | $p_x$ | $p_y$ | $p_z$ |
| spin-up | $p_x$ | | $i\lambda_{OS}$ | | | | $-\lambda_{OS}$ |
| | $p_y$ | $-i\lambda_{OS}$ | | | | | $i\lambda_{OS}$ |
| | $p_z$ | | | | $\lambda_{OS}$ | $-i\lambda_{OS}$ | |
| spin-down | $p_x$ | | | $\lambda_{OS}$ | | $-i\lambda_{OS}$ | |
| | $p_y$ | | | $i\lambda_{OS}$ | $i\lambda_{OS}$ | | |
| | $p_z$ | $-\lambda_{OS}$ | $-i\lambda_{OS}$ | | | | |

| Table 3 Band Gaps of Monolayer of VA Elements | | N | P | As | Sb | Bi |
|---|---|---|---|---|---|---|
| Without SOC | Band gap/eV | 3.70 | 1.97 | 1.59 | 1.26 | 0.55 |
| | Gap type | Indirect | Indirect | Indirect | Indirect | Direct |
| | VBM | Not Γ | Not Γ | Γ | Γ | Γ |
| | CBM | Not Γ | Not Γ | Not Γ | Not Γ | Γ |
| With SOC | Band gap/eV | 3.70 | 1.97 | 1.81 | 1.00 | 0.43 |
| | Gap type | Indirect | Indirect | Indirect | Indirect | Indirect |
| | VBM | Not Γ | Not Γ | Γ | Γ | Not Γ |
| | CBM | Not Γ | Not Γ | Not Γ | Not Γ | Γ |
| Change of band gap due to SOC/eV | | 0.00 | 0.01 | 0.21 | -0.26 | -0.12 |

| Table 4 Tight-Binding Parameters for Monolayer of VA Elements (eV) | | | | | | | | | |
|---|---|---|---|---|---|---|---|---|---|
| Element | N | P | | As | | Sb | | Bi | |
| | | No SOC | SOC | No SOC | SOC | No SOC | SOC | No SOC | SOC |
| $\varepsilon_s$ | -6.80 | -5.02 | -5.02 | -4.90 | -7.53 | -5.49 | -6.79 | -7.09 | -8.89 |
| $\varepsilon_p$ | -1.67 | -0.565 | -0.565 | -0.275 | -0.452 | -0.252 | -0.613 | -0.105 | -0.563 |
| $\gamma_{ss}$ | -1.49 | -1.07 | -1.08 | -0.351 | -1.22 | -0.448 | -0.985 | -0.0515 | -0.864 |
| $\gamma_{sp}$ | 4.76 | 2.89 | 2.89 | 1.98 | 2.72 | 1.85 | 2.20 | 1.61 | 1.99 |
| $\gamma_{pp\sigma}$ | -5.94 | -3.34 | -3.34 | -2.91 | -2.77 | -2.29 | -2.26 | -2.04 | -2.02 |
| $\gamma_{pp\pi}$ | 1.56 | 0.767 | 0.767 | 0.684 | 0.606 | 0.557 | 0.472 | 0.394 | 0.395 |
| $\gamma'_{ss}$ | 0.216 | 0.260 | 0.261 | 0.0986 | -0.0529 | 0.0490 | -0.0735 | -0.0186 | -0.124 |
| $\gamma'_{sp}$ | -0.0235 | -0.0562 | -0.0562 | -0.0637 | 0.179 | -0.0384 | 0.0280 | -0.0200 | -0.124 |
| $\gamma'_{pp\sigma}$ | 1.38 | 1.15 | 1.15 | 0.858 | 0.786 | 0.653 | 0.646 | 0.535 | 0.551 |
| $\gamma'_{pp\pi}$ | -0.499 | -0.377 | -0.377 | -0.233 | -0.174 | -0.162 | -0.130 | -0.109 | -0.140 |
| $\lambda_I/10^{-3}$ | — | — | -0.141 | — | 0 | — | 1.97 | — | 0 |
| $\lambda_R/10^{-2}$ | — | — | 0.406 | — | 0 | — | 0.278 | — | 0 |
| $\lambda_{OS}$ | — | — | -0.0106 | — | -0.285 | — | -0.299 | — | -0.630 |

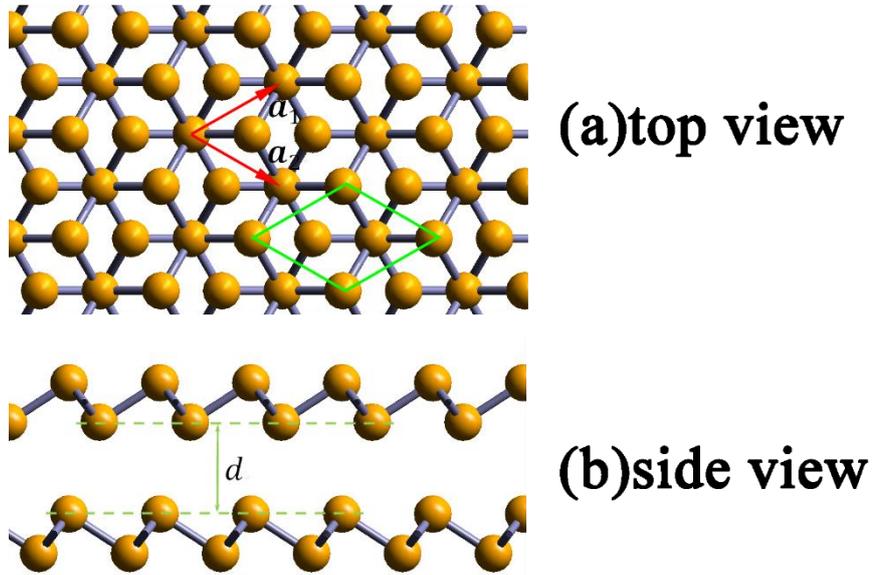

Figure 1 (a) Top view and (b) Side view of three-dimensional layered structure of VA elements. Red arrows are the lattice vectors in the plane. Green rhombus represents the unit cell. $d$ is the inter-layer distance.

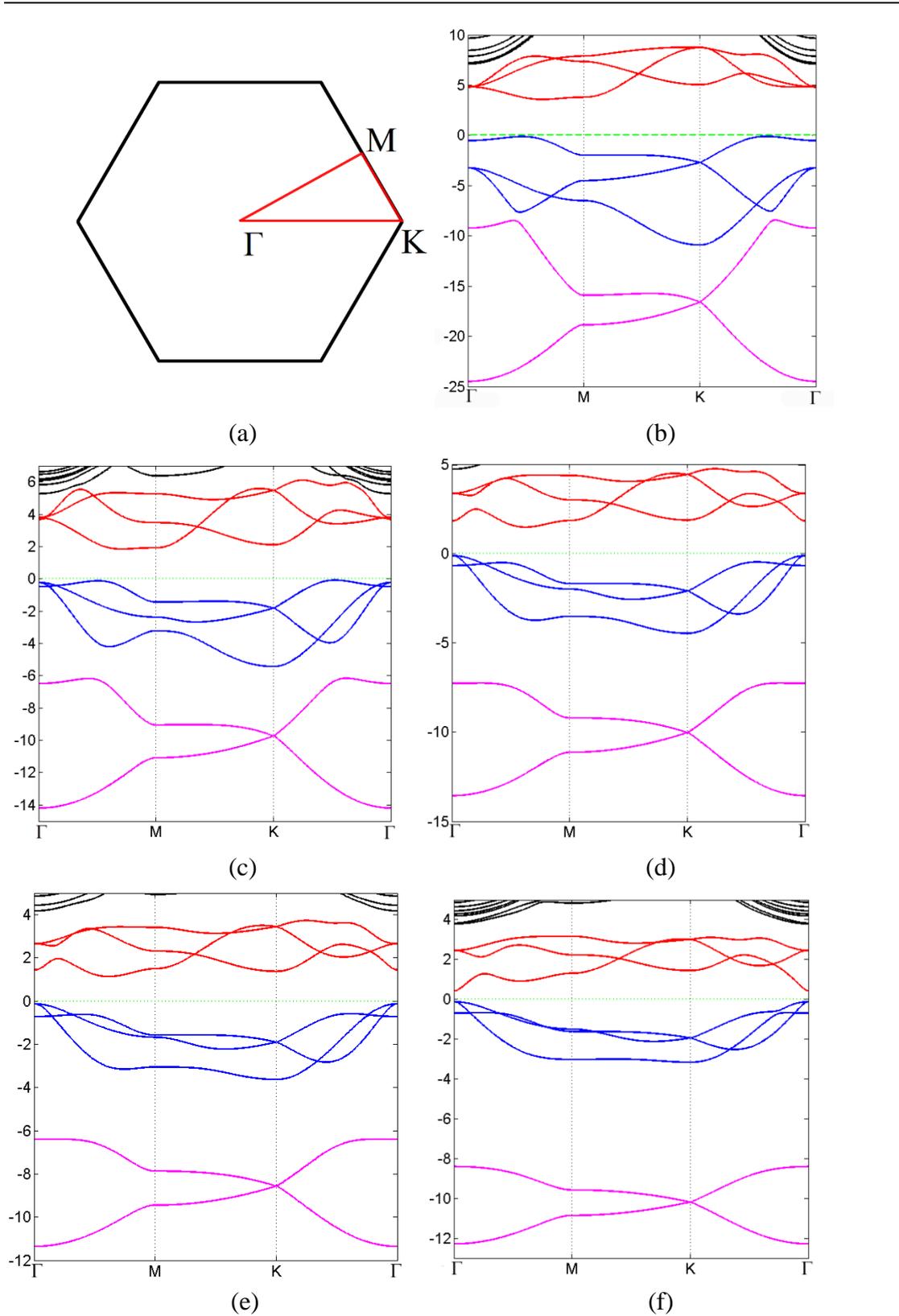

Figure 2 (a) The first Brillouin zone and the high symmetry points of two-dimensional honeycomb lattice. (b)-(f) Band structure evolution of 2D monolayer of VA elements as it goes from nitrogen to bismuth. Different colors refer to different branches of bands. Fermi energy is set to zero denoted by green dash line.

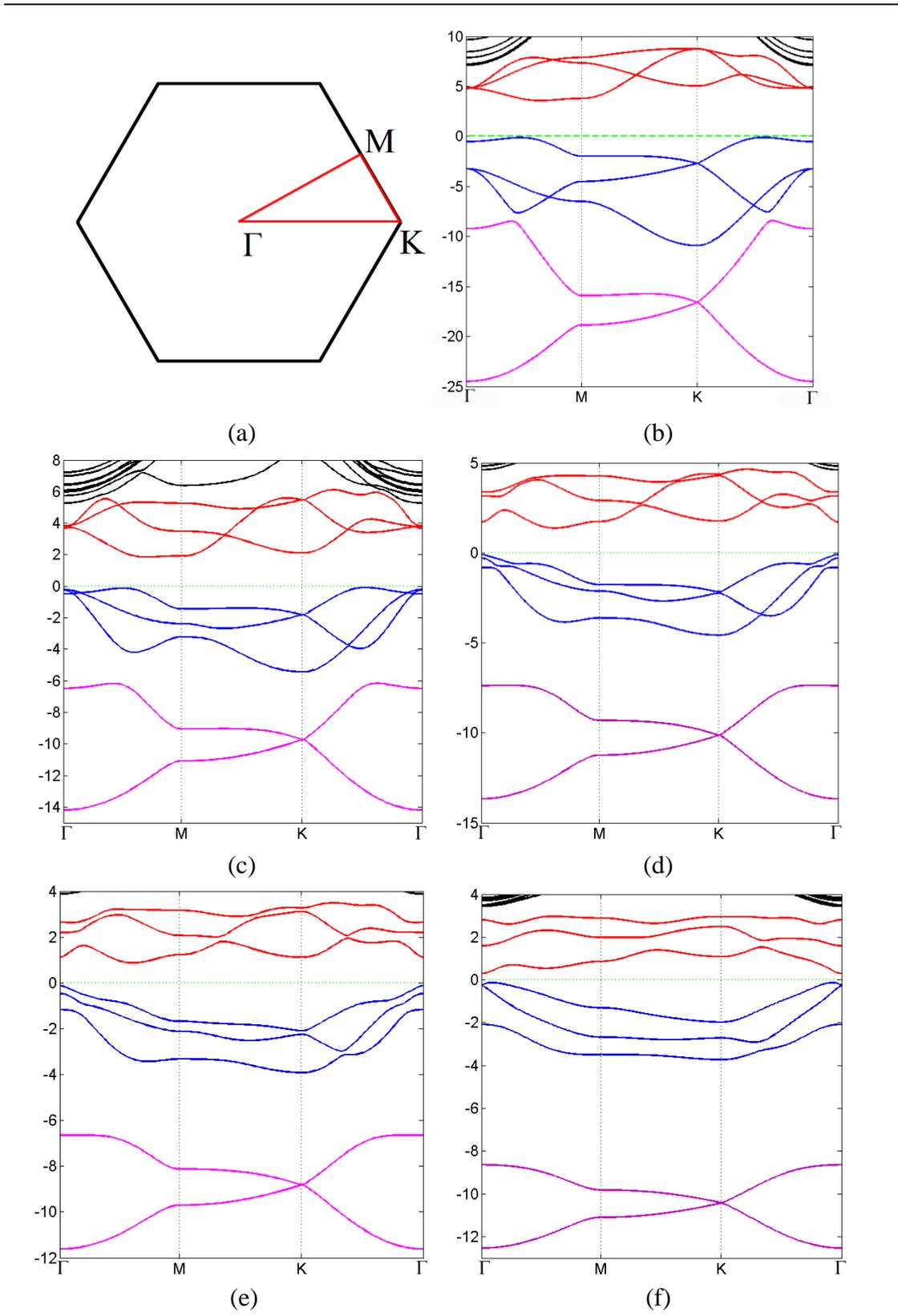

Figure 3 The same as Figure 2, but with SOC.

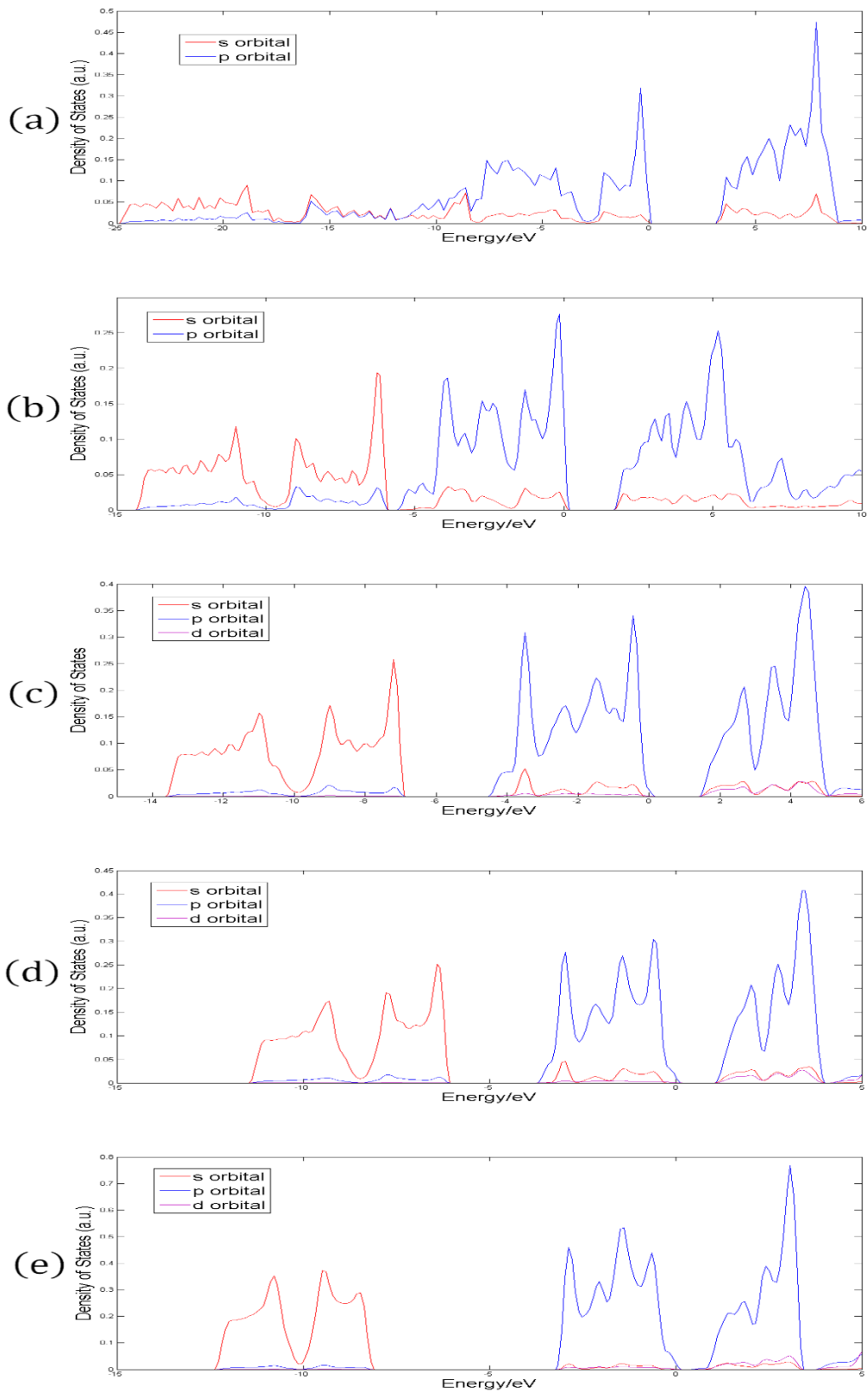

Figure 4 The density of states of monolayers of (a) nigrogen, (b) phosphorus, (c) arsenic, (d) antimony and (e) bismuth. Red, blue and purple solid lines denote the projection into s orbital, p orbital and d orbital, respectively. Fermi energy is set to zero.

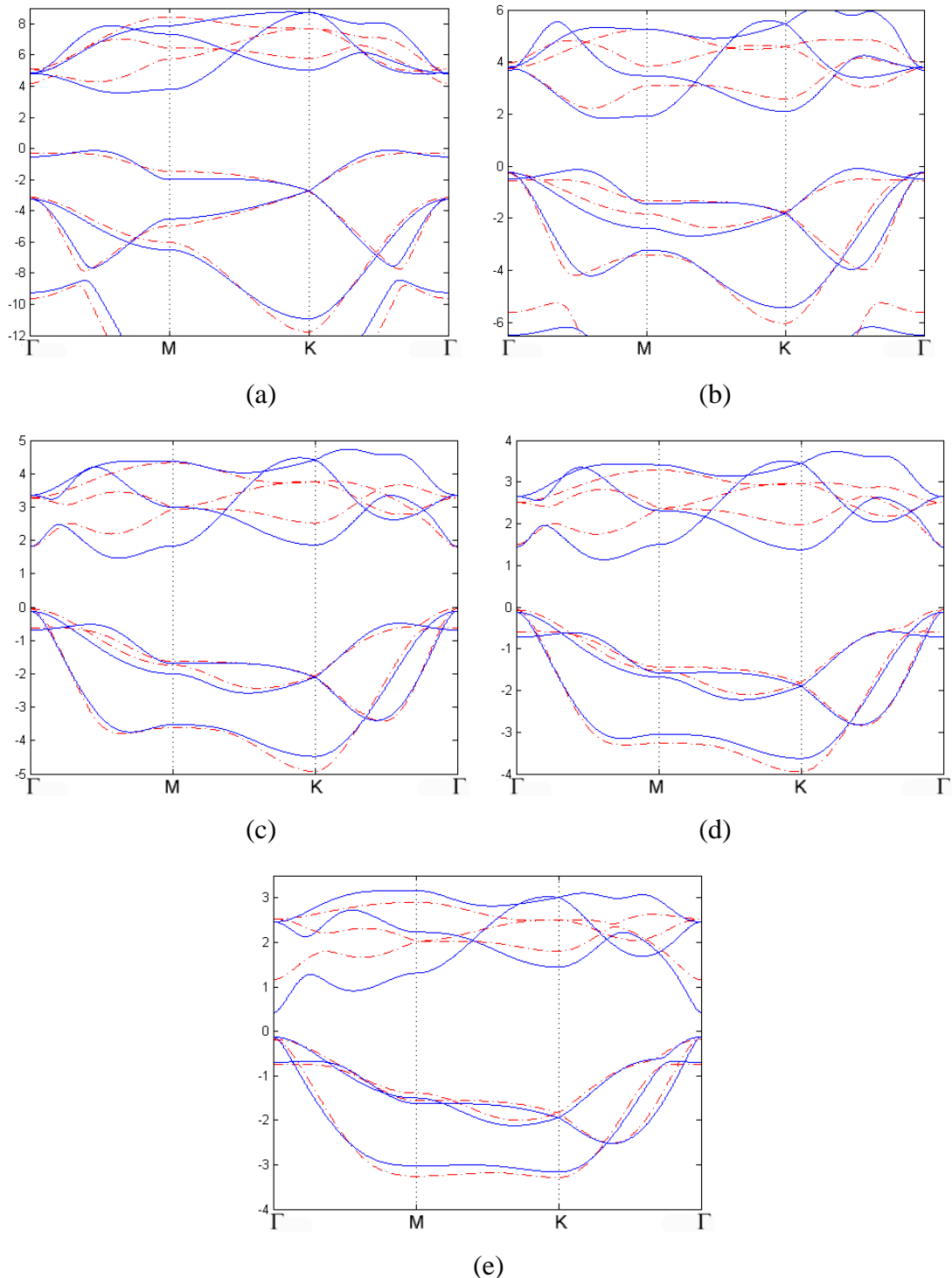

Figure 5 (a)-(e) Tight-binding band structures of 2D monolayer of VA elements as it goes from nitrogen to bismuth (red dotted line), compared with first principal results (blue solid line). The first Brillouin Zone and the high symmetric points are the same as two previous figures.

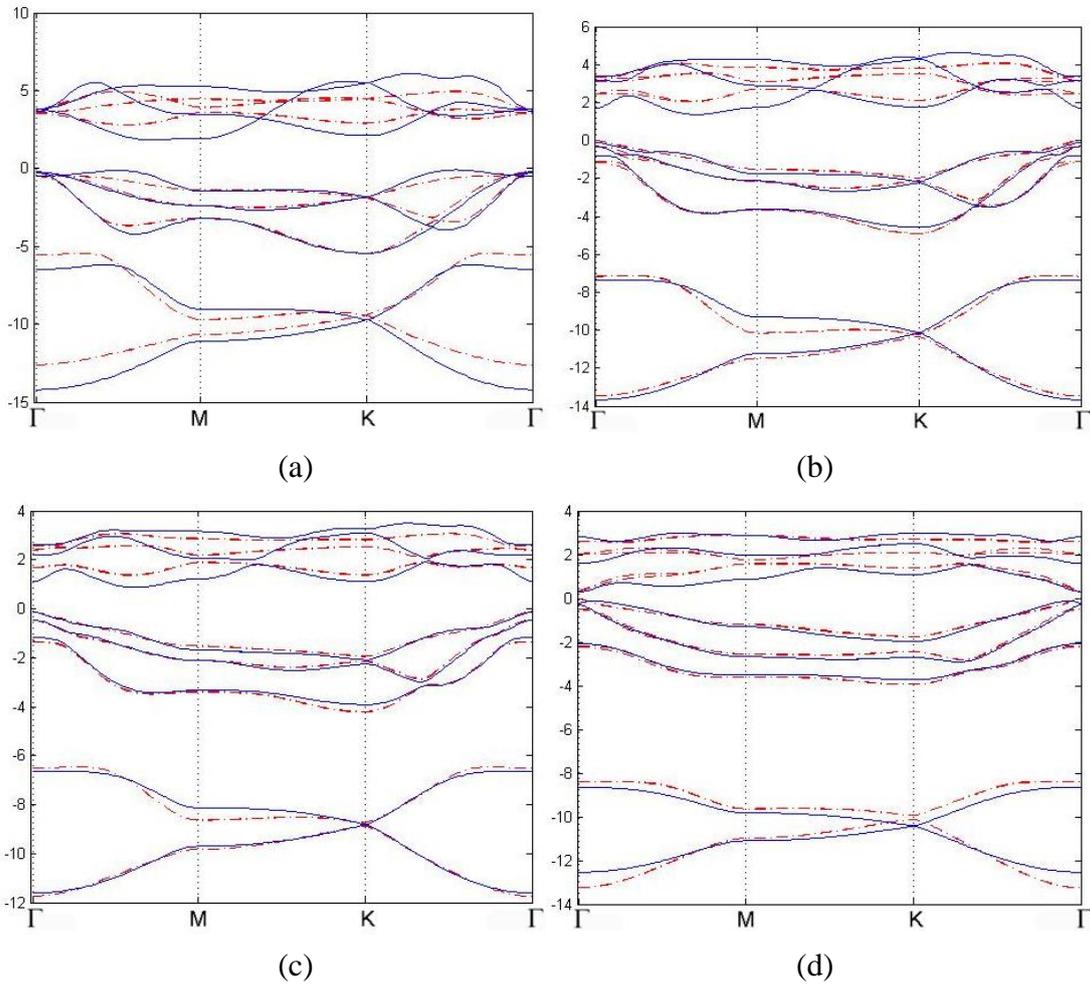

Figure 6 (a)-(d) Tight-binding band structures of 2D monolayer of VA elements as it goes from phosphorus to bismuth (red dotted line). All others are the same as Figure 5, but with SOC.

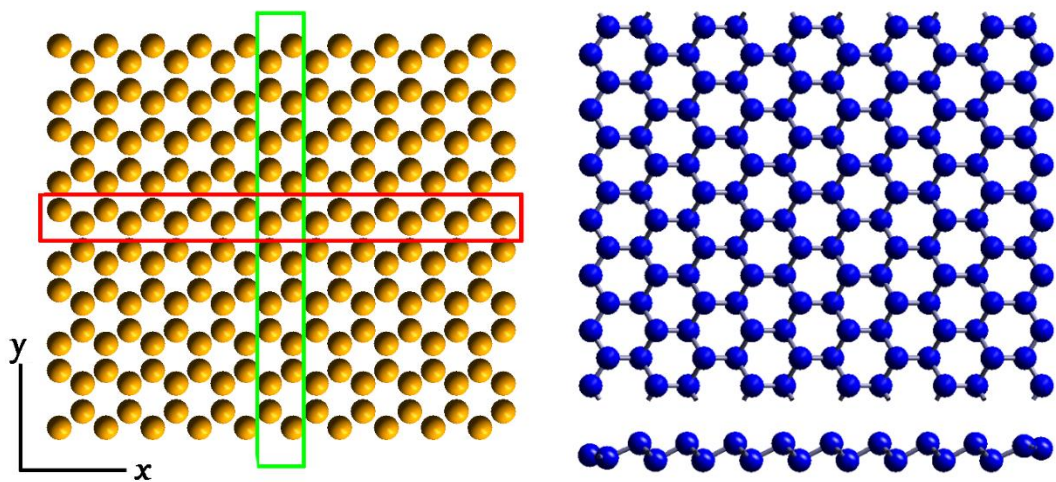

(a) (b)

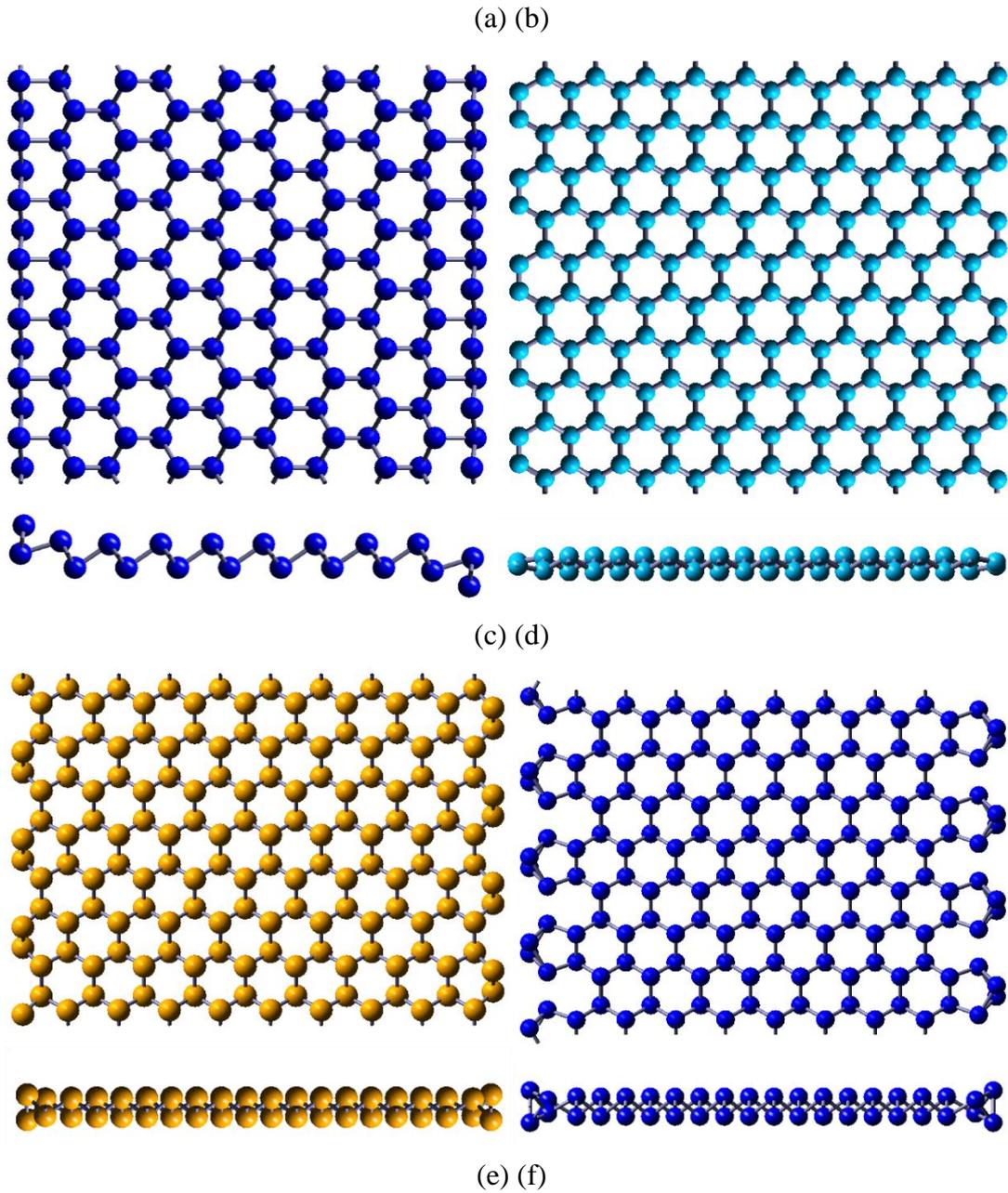

(c) (d)

(e) (f)

Figure 7(a) ZNR and ANR with 20 atoms wide. The system is periodic in x direction for ZNR and in y direction for ANR. The green and red rectangles represent the atoms in the unit cell for ZNR and ANR, respectively. Vacua of 15 Å are inserted on both sides of the nanoribbon. (b) and (c) are the optimized structure of ZNR for Nitrogen and Bismuth, respectively. (d), (e) and (f) are the optimized structure of ANR Nitrogen, Phosphorus, and Bismuth, respectively.

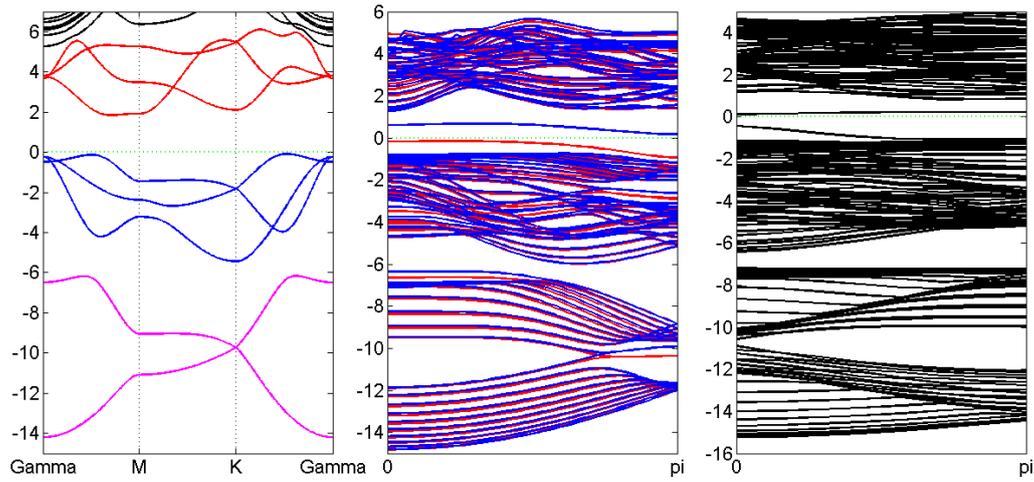

(a) (b) (c)

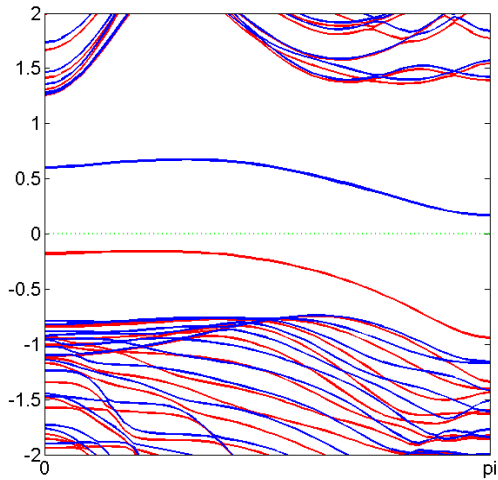
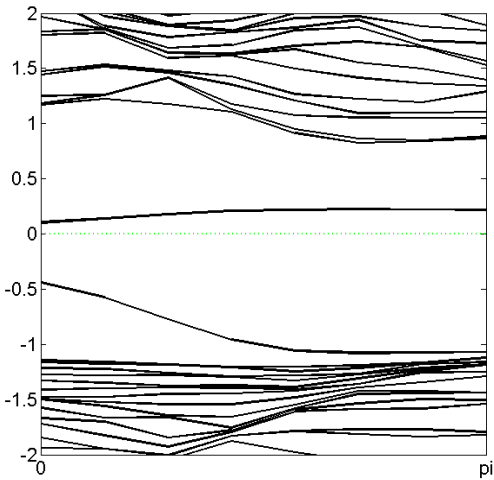

(d) (e)

Figure 8 Band structure of zigzag blue phosphorus ZNR (b) and ANR (c) compared with that of monolayer (a). (d) and (e) Same as (b) and (c), but the energy scale is different. The Fermi level is denoted by green dash line. Red and blue solid lines in (b) and (d) refer to spin-up and spin-down, respectively.

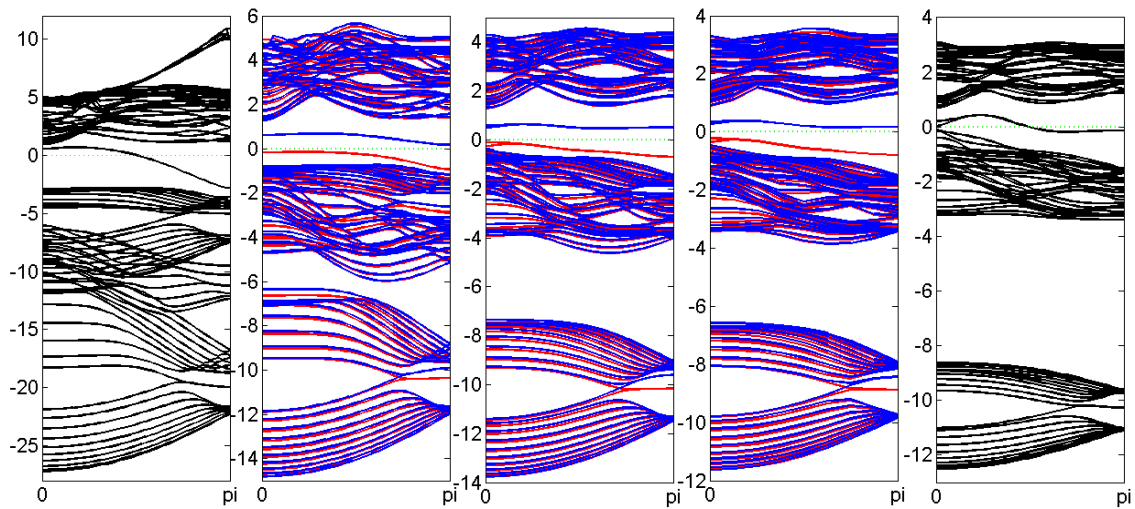

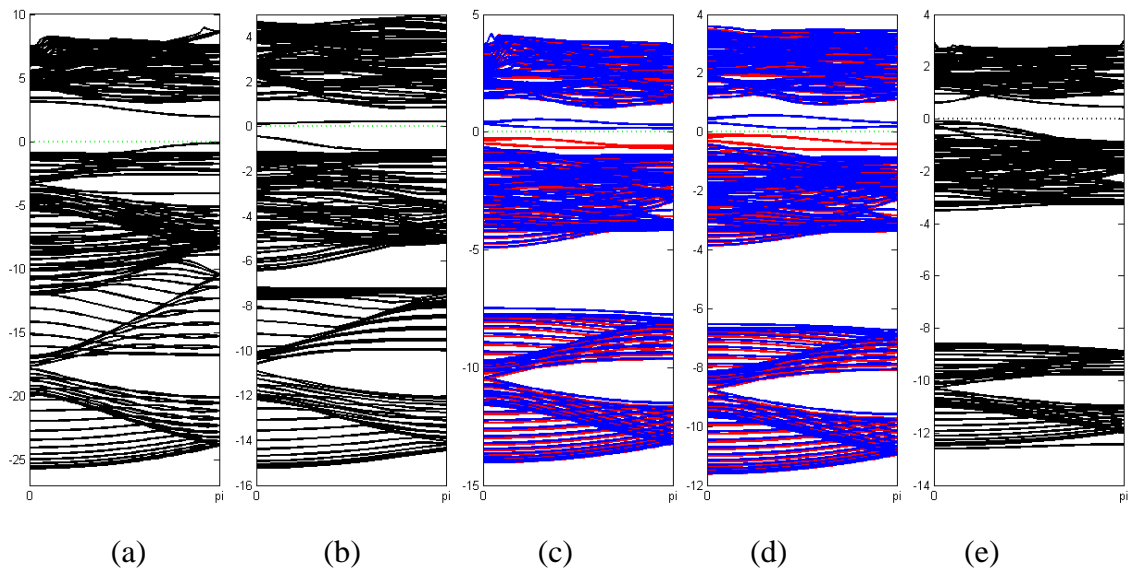

(a) (b) (c) (d) (e)

Figure 9 Band structure of ZNR (upper panel) and ANR (lower panel) for (a) nitrogen, (b) phosphorus, (c) arsenic, (d) antimony and (e) bismuth, respectively. The Fermi level is denoted by green dash line. For (b), (c) and (d), Red and blue solid lines refer to spin-up and spin-down, respectively.

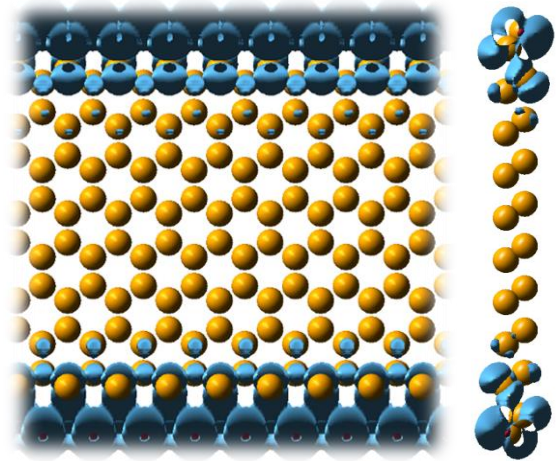

Figure 10 Charge density distribution of edge state in ZNR of P.